\documentclass[12pt,a4paper]{article}
\usepackage{amsmath,amssymb,bm,ascmac,bbm,mathtools,here}
\usepackage[dvipdfmx, usenames]{color}
\usepackage[dvipdfmx]{graphicx}
\usepackage{xcolor}
\usepackage{here}
\usepackage{authblk}

\newcommand{\tr}{\text{tr}}

\newcommand{\mcal}{\mathcal}
\newcommand{\mbb}{\mathbb}

\newcommand{\la}{\langle}
\newcommand{\ra}{\rangle}

\newcommand{\vep}{\varepsilon}

\begin{document}
\title{Symmetry enhancement in RCFT}
\author{Ken KIKUCHI}
\affil{Yau Mathematical Sciences Center, Tsinghua University, Beijing, China}
\date{}
\maketitle

\begin{abstract}
We propose when and why symmetry enhancements happen in massless renormalization group (RG) flows to two-dimensional rational conformal field theories (RCFTs). We test our proposal against known RG flows from unitary minimal models. We also suggest which sign of the relevant coupling triggers the massless RG flow. The other sign triggers massive RG flows to topological quantum field theories (TQFTs). We comment on their ground state degeneracies.
\end{abstract}


\makeatletter
\renewcommand{\theequation}
{\arabic{section}.\arabic{equation}}
\@addtoreset{equation}{section}
\makeatother

\section{Introduction}
Typical problems in physics are given as follows; problems are defined at ultraviolet (UV), and answers are their infrared (IR) behaviors. A famous example is the quantum chromodynamics (QCD). Its UV description has been known for decades. However, its IR behavior remained uncertain for a long time. Although it was believed spontaneous (chiral) symmetry breaking (SSB) and confinement take place, strong coupling in IR has prevented us from proving them exactly.\footnote{See, say \cite{M,CGMT}, for recent remarkable developments in this direction.} As illustrated by this example, we are usually interested in the fate of symmetries in IR.

Contrary to SSB, sometimes symmetries enhance at long distances. One canonical example is this; we usually define a theory on a lattice in UV, and hope it is described by continuum quantum field theory (QFT) in IR. Namely, we believe the discrete translation and rotation symmetries in the UV enhance to much larger continuous Poincar\'e symmetry in the IR. Some other influential examples are four-dimensional UV $\mcal N=1$ supersymmetry enhancing to $\mcal N=2$ supersymmetry in IR (spacetime symmetry enhancement) \cite{MS}, and $SU(8)$ UV symmetry enhancing to $E_7$ IR symmetry (flavor symmetry enhancement) \cite{DG}.

As one can feel from these examples, the phenomenon remains mysterious. In particular, we lack an understanding of when and why symmetry enhances. In this paper, we would like to address this point. Our proposal can explain when and why the symmetry enhancements should happen.

Before stating our proposal, let us recall what a symmetry is. Recently, the notion of symmetry was generalized \cite{GKSW}. Symmetries are generated by charges. Classically, a charge is defined on a time slice, and it is preserved under time translation of the slice. More generally, these two properties can be rephrased as follows; (ordinary) symmetry charges are defined on codimension-one defects which are topological. Using these two axioms, \cite{GKSW} generalized ordinary zero-form symmetries to higher-form symmetries by increasing codimensions of topological defects. The generalized global symmetries gave, for example, unified description of de/confinement phase transition as an SSB of one-form symmetry. There is another direction of generalization; note that the axioms do not require the symmetries to be invertible. Namely, symmetries can fail to form groups. In other words, monoid-like symmetries are allowed. These monoid-like symmetries are sometimes called non-invertible symmetries.\footnote{Condensed matter physicists also call them algebraic symmetries (see for example \cite{JW}) because products of non-invertible topological defects in general give a sum of topological defects.} They are also called category symmetries because they are described by certain categories.

How can we explain enhancements of category symmetries? We focus on two-dimensional rational conformal field theories (RCFTs). Zero-form symmetries in two dimensions are generated by topological defect lines (TDLs). They in general form fusion category.\footnote{See e.g. \cite{TW1,TW2,KORS} for recent applications of fusion category symmetry to two-dimensional QFTs.} A category symmetry of an RCFT with enlarged chiral algebra $\mcal V$ is described by modular tensor category (MTC) $\text{Rep}(\mcal V)$ \cite{MS1,MS2}, or modular category for short. Therefore, in massless RG flows to IR RCFTs, symmetries in IR should be described by MTCs. By requiring this (with the help of $c$-theorem \cite{cthm}), we can correctly explain symmetry enhancements in RCFTs. Namely, when a putative symmetry category in IR is either non-modular or inconsistent with the $c$-theorem, the category $\mcal C$ should be enhanced to a larger category $\mcal D\supset\mcal C$ to make it both modular and consistent with the $c$-theorem.\footnote{Two remarks: firstly, we are assuming IR theories to be RCFTs in this paper. For other RG flows, say, to irrational CFTs, our proposal does not claim anything. Secondly, although our proposal can explain when symmetry enhancements should happen, it cannot explain \textit{how large} $\mcal D$ should be. We leave these important points for future.} To check our proposal, we study known RG flows from unitary minimal models.

To motivate our proposal, let us briefly look at two examples, one trivial and one non-trivial. We begin from the trivial one. Let us take the tricritical Ising model as a UV theory. The theory has six TDLs, among which only three TDLs survive a relevant deformation with $\phi_{1,3}$. (The details are explained in the next section.) It turns out that the surviving TDLs, which a priori form just a braided fusion category (BFC), actually form a rank\footnote{The `size' of fusion categories are called rank. More precisely, it counts the number of isomorphism classes of simple objects.} three MTC. Since the putative rank three MTC can have central charge smaller than that in UV, our proposal does not require symmetry enhancement, consistent with known massless RG flow to the critical Ising model. Next, let us look at an example with symmetry enhancement. We take tetracritical Ising model as a UV theory. The theory has 10 TDLs, among which only four survive the $\phi_{1,3}$-relevant deformation. One finds the rank four BFC is $\textit{not}$ an MTC. One way out is to add extra TDLs to make it modular. Can an additional TDL make it modular? No. There is no rank five MTC containing the surviving BFC. So at least two additional TDLs should emerge to satisfy our requirements. In fact, there is only one rank six MTC containing the rank four BFC. Furthermore, the candidate can have central charge smaller than that in UV. A natural minimal scenario\footnote{This raises a new notion of ``naturality"; it is unnatural that symmetries enhance without any reason. For example, it is unlikely that a symmetry with, say, rank three enhances to rank 100 without underlying mechanism. Our proposal makes symmetry enhancements we study natural.} with just two emergent TDLs is consistent with known RG flow to the tricritical Ising model. These observations strongly suggest symmetry enhancements happen when surviving BFC is non-modular or surviving MTC cannot have central charge smaller than that in UV. (We will see the second requirement in action later.) In the next section (and in some appendices), we study these examples in more detail. We also generalize the consideration to all minimal models $M(m+1,m)$ with $m\ge3$.

Our notations in minimal models are summarized in Appendix \ref{notation}. Tools and some facts about BFCs we employ are summarized in Appendix \ref{MTC}. The case of massive RG flow is considered in Appendix \ref{massive}.

\section{Tests: Massless RG flow from unitary minimal model}
Let us expose our proposal to tests in this section. In order to check it, we focus on RG flows where IR theories are known. More concretely, we study RG flows from unitary minimal models $M(m+1,m)$ (with $m\ge3$) triggered by a relevant operator $\phi_{1,3}$. For $m>3$, it is known that one sign of the relevant coupling $\lambda_{1,3}$ triggers a massless RG flow to the next minimal model $M(m,m-1)$, and the other sign triggers a massive RG flow to a TQFT possibly with some ground state degeneracies. In the Appendix \ref{massive}, we fix which sign leads to massless RG flows (so the other sign leads to massive RG flows). In the latter case, we also comment on ground state degeneracies.

\subsection{Generality}\label{general}
Let us consider generic cases with $m>3$. The UV minimal model $M(m+1,m)$ has
\[ \frac12m(m-1) \]
TDLs. Perturbation of the theory with the relevant operator $\phi_{1,3}$ breaks some of them. Here, we can prove the following lemma (see Appendix \ref{proof}):\\\newline
\textbf{Lemma} \textit{Only $(m-1)$ TDLs $\mcal L_{r,1}$ with $r=1,2,\dots,m-1$ survive the relevant $\phi_{1,3}$-deformation (this result is also true for $m=3$).} $\square$\\\newline
When $m>3$, it is known that the IR theory is the next minimal model $M(m,m-1)$ with
\[ \frac12(m-1)(m-2) \]
TDLs. So the difference
\begin{equation}
\frac12(m-1)(m-2)-(m-1)=\frac12(m-1)(m-4)\label{difference}
\end{equation}
should be compensated somehow (we will comment on this point later).

Now, let us also match TDLs in the two theories. As demonstrated by the lemma, the surviving TDLs have Kac indices $(r,1)$ with $r=1,2,\dots,m-1$. Thus their quantum dimensions are given by
\begin{equation}
d_{(r,1)}^\text{UV}=\frac{\sin\frac{\pi r}m}{\sin\frac\pi m}.\label{survqd}
\end{equation}
These quantum dimensions should be matched by TDLs in the IR theory. The theory has one smaller $m$. So its TDLs have quantum dimensions
\begin{equation}
d_{(t,u)}^\text{IR}=\frac{\sin\frac{\pi t}{m-1}\sin\frac{\pi u}m}{\sin\frac\pi{m-1}\sin\frac\pi m}.\label{IRqd}
\end{equation}
We find the quantum dimensions are matched as\footnote{Just matching the quantum dimensions, there is a $\mbb Z_2$ ambiguity caused by fusion with the $\mbb Z_2$ TDL. In the previous version, we were careless about this point, but taking into account the special role played by $\mcal L_{2,1}$ and $\mcal L_{1,2}=\mcal L_{m-1,m-1}$ under fusion, we can fix the ambiguity.}
\begin{equation}
d_{(1,1)}^\text{UV}=d_{(1,1)}^\text{IR},\quad d_{(r,1)}^\text{UV}=d_{(m-2,m-r)}^\text{IR},\label{qdmatching}
\end{equation}
where $r\neq1$. Therefore, we can identify TDLs in two theories as follows:
\begin{equation}
    \begin{array}{cccccc}
    \text{UV}:&\mcal L_{1,1}&\mcal L_{2,1}&\cdots&\mcal L_{m-2,1}&\mcal L_{m-1,1}\\
    &\downarrow&\downarrow&\cdots&\downarrow&\downarrow\\
    \text{IR}:&\mcal L_{1,1}&\mcal L_{m-2,m-2}&\cdots&\mcal L_{m-2,2}&\mcal L_{m-2,1}
    \end{array}.\label{matchingTDLs}
\end{equation}
By connecting the two neighboring minimal models successively, one can in principle reach an arbitrary minimal model $M(m+1,m)$ from another $M(m'+1,m')$ with $m'>m$.

Finally, let us briefly comment on odd and even $m$ cases in turn.

\subsubsection{Odd $m$}
We find symmetry enhancement is needed when $m$ (larger than three) is odd (see the Appendix \ref{braid}). In this case $m=2M+1$ ($M=2,3,\dots$), our proposal claims the difference (\ref{difference}) should be compensated by emergent TDLs which are required to make the surviving BFC to an MTC. Therefore, we get two-fold statements in this case:
\begin{itemize}
\item Assuming the knowledge of IR theories, we obtain conjectures on classification of MTCs; there is no MTC satisfying three conditions simultaneously; (i) whose rank $r$ is in the range $(m-1)<r<\frac12(m-1)(m-2)$; (ii) whose central charge $c$ obey $c<c_\text{UV}=1-\frac6{m(m+1)}$; (iii) which contains the fusion relations of the surviving BFC with rank $(m-1)$.
\item If we only know an IR theory should be some RCFT,\footnote{So in this case, we cannot use (\ref{difference}), nor (\ref{IRqd}) to (\ref{matchingTDLs}). This situation would occur, for example, when anomalies forbid massive phases.} but with the classification of MTCs at hand, one can \textit{predict} symmetry enhancement by looking for an MTC consistent with the $c$-theorem and which enlarges the surviving BFC as we did in the Introduction.
\end{itemize}
If our proposal requires symmetry enhancement, we claim the two statements to hold also for even $m$.

\subsubsection{Even $m$}
In this case $m=2M$ ($M=2,3,\dots$), the requirement of modularity is automatically satisfied as we show in Appendix \ref{braid}. So we only have to compare central charges of putative MTCs and $c_\text{UV}$. In fact, we will see consistency with the $c$-theorem requires symmetry enhancement at $m=6$. Although we could not check our proposal in $m\ge8$ due to the lack of classification of MTCs, we believe the two requirements $-$ modularity and consistency with the $c$-theorem $-$ explain symmetry enhancements in two-dimensional RCFTs.

\subsection{Examples}
Let us study examples with small $m$ in more details.
\begin{itemize}
    \item $m=4$: Tricritical Ising model\\
    In this case, the Kac indices are given by
    \[ E_{5,4}=\{(1,1),(2,1),(2,2),(3,1),(3,2),(3,3)\}. \]
    Among these, only three TDLs with
    \[ \widetilde E_{5,4}=\{(1,1),(2,1),(3,1)\} \]
    survive the relevant $\phi_{1,3}$-deformation. As we showed in the Appendix \ref{braid}, the symmetric centralizer of the surviving BFC $\mcal C$ is trivial.\footnote{Explicit computation gives the braiding matrix
    \[ \lambda=\begin{pmatrix}1&1&1\\1&*&-1\\1&-1&1\end{pmatrix} \]
    in the basis $\{\mcal L_{1,1},\mcal L_{2,1},\mcal L_{3,1}\}$. It is clear that there is no nontrivial transparent TDL.} So the BFC is automatically an MTC. In fact, fusion relations of $\mcal C$ coincide with the one in \cite{rank6} with $SU(2)_2$ realization. Since the MTC has central charge $\frac{2n+1}2$ ($n<4$), which can be smaller than that in UV $c_\text{UV}=\frac7{10}$, our requirements are satisfied by the candidate with $n=0$, or $c=\frac12$. Hence, symmetry enhancement is not necessary, consistent with the known RG flow to the critical Ising model with $c_\text{IR}=\frac12<\frac7{10}=c_\text{UV}$:
    \begin{equation}
    \begin{array}{ccccc}
    \text{UV}:&\mcal L_{1,1}&\mcal L_{2,1}&\mcal L_{3,1}\\
    &\downarrow&\downarrow&\downarrow\\
    \text{IR}:&\mcal L_{1,1}&\mcal L_{2,2}&\mcal L_{2,1}
    \end{array}.\label{matchingtriTDLs'}
    \end{equation}
    \item $m=5$: Tetracritical Ising model\\
    In this case, the Kac indices are given by
    \[ E_{6,5}=\{(1,1),(2,1),(2,2),(3,1),(3,2),(3,3),(4,1),(4,2),(4,3),(4,4)\}. \]
    Among these, only four TDLs with
    \[ \widetilde E_{6,5}=\{(1,1),(2,1),(3,1),(4,1)\} \]
    survive the relevant $\phi_{1,3}$-deformation. Now, $(r,s)=(4,1)$ gives $\lambda=1$ for all Kac indices in $\widetilde E_{6,5}$.\footnote{The full braiding matrix is given by
    \[ \lambda=\begin{pmatrix}1&1&1&1\\1&*&*&1\\1&*&*&1\\1&1&1&1\end{pmatrix}, \]
    in the basis $\{(1,1),(2,1),(3,1),(4,1)\}$.} Hence, the symmetry should enhance to make the rank four BFC $\mcal C$ an MTC. In the math literature, MTCs up to rank five have been classified in \cite{rank5}. One finds none of them contains the fusion relations of $\mcal C$. Therefore, at least two TDLs should emerge to satisfy our requirements. In fact, there is only one rank six MTC\footnote{The one with a realization $SU(2)_3/\mbb Z_2\times SU(2)_2$. One can see the rank six MTC contains fusion relations of the surviving BFC with identifications $\phi_{1,1}^\text{here}=1^\text{there},\phi_{2,1}^\text{here}=\phi_4^\text{there},\phi_{3,1}^\text{here}=\phi_1^\text{there},\phi_{4,1}^\text{here}=\phi_2^\text{there}$.} \cite{rank6} enlarging $\mcal C$. According to the paper, it has central charge of the form $c=\frac{2n+1}{10}$, which can be smaller than  $c_\text{UV}=\frac45$. This is consistent with the known RG flow to the tricritical Ising model (with $n=3$, or $c_\text{IR}=\frac7{10}$). The TDLs are matched as follows:
    \begin{equation}
    \begin{array}{ccccc}
    \text{UV}:&\mcal L_{1,1}&\mcal L_{2,1}&\mcal L_{3,1}&\mcal L_{4,1}\\
    &\downarrow&\downarrow&\downarrow&\downarrow\\
    \text{IR}:&\mcal L_{1,1}&\mcal L_{3,3}&\mcal L_{3,2}&\mcal L_{3,1}
    \end{array}.\label{matchingtetraTDLs'}
    \end{equation}
    The other two TDLs
    \[ \{\mcal L_{2,1},\mcal L_{2,2}\} \]
    in IR are emergent.
    \item $m=6$: Pentacritical Ising model\\
    In this case, the Kac indices are given by
    \begin{align*}
    E_{7,6}=\{&(1,1),(2,1),(2,2),(3,1),(3,2),(3,3),(4,1),(4,2),(4,3),(4,4),\\
    &(5,1),(5,2),(5,3),(5,4),(5,5)\}.
    \end{align*}
    Among these, only five TDLs with
    \[ \widetilde E_{7,6}=\{(1,1),(2,1),(3,1),(4,1),(5,1)\} \]
    survive the relevant $\phi_{1,3}$-deformation. As we showed in the Appendix \ref{braid}, the symmetric centralizer is trivial. Hence, the surviving rank five BFC $\mcal C$ is actually an MTC. In fact, the fusion relations of $\mcal C$ 
    \begin{table}[H]
    \begin{center}
    \begin{tabular}{c|c|c|c|c|c}
	&$\mcal L_{1,1}$&$\mcal L_{2,1}$&$\mcal L_{3,1}$&$\mcal L_{4,1}$&$\mcal L_{5,1}$\\\hline
	$\mcal L_{1,1}$&$\mcal L_{1,1}$&$\mcal L_{2,1}$&$\mcal L_{3,1}$&$\mcal L_{4,1}$&$\mcal L_{5,1}$\\\hline
	$\mcal L_{2,1}$&&$\mcal L_{1,1}+\mcal L_{3,1}$&$\mcal L_{2,1}+\mcal L_{4,1}$&$\mcal L_{3,1}+\mcal L_{5,1}$&$\mcal L_{4,1}$\\\hline
	$\mcal L_{3,1}$&&&$\mcal L_{1,1}+\mcal L_{3,1}+\mcal L_{5,1}$&$\mcal L_{2,1}+\mcal L_{4,1}$&$\mcal L_{3,1}$\\\hline
	$\mcal L_{4,1}$&&&&$\mcal L_{1,1}+\mcal L_{3,1}$&$\mcal L_{2,1}$\\\hline
	$\mcal L_{5,1}$&&&&&$\mcal L_{1,1}$
    \end{tabular}
    \end{center}
    \end{table}
    coincide with the one in \cite{rank6}.\footnote{The one with a realization $SU(2)_4$ under identifications
    \[ \phi_{1,1}^\text{here}=1^\text{there},\quad\phi_{2,1}^\text{here}=\phi_1^\text{there},\quad\phi_{3,1}^\text{here}=\phi_2^\text{there},\quad\phi_{4,1}^\text{here}=\phi_3^\text{there},\quad\phi_{5,1}^\text{here}=\phi_5^\text{there}, \]
    or
    \[ \phi_{1,1}^\text{here}=1^\text{there},\quad\phi_{2,1}^\text{here}=\phi_3^\text{there},\quad\phi_{3,1}^\text{here}=\phi_2^\text{there},\quad\phi_{4,1}^\text{here}=\phi_1^\text{there},\quad\phi_{5,1}^\text{here}=\phi_5^\text{there}. \]
    Two possibilities appear due to the invariance of fusion relations under $\mcal L_{2,1}\leftrightarrow\mcal L_{4,1}$. In the $SU(2)_4$ language, the transformation is realized by the $\mbb Z_2$ outer automorphism acting on affine weights $[3;1]\leftrightarrow[1;3]$.} However, according to the paper, the rank five MTC has central charge $c=2$ mod 4. Assuming the unitarity, this means $c=2,6,10,\dots$. Thus, it cannot be smaller than $c_\text{UV}=\frac67$. One way out to remedy the problem is to enlarge the MTC $\mcal C$ to another MTC $\mcal D\supset\mcal C$ so that central charge can be smaller than in UV. How large should $\mcal D$ be? Unfortunately, we cannot answer this question\footnote{At least we checked there is no rank six MTC in the list \cite{rank6} enlarging $\mcal C$. We also checked none of the MTCs in partial classification \cite{rank9} up to $r=9$ satisfy the conditions. This suggests the IR theory would have $r\ge10$.} because higher rank MTCs have not been completely classified yet to the best of our knowledge. However, we believe the smallest possible MTC accommodating central charge $c=\frac45$ would have rank 10, i.e., the MTC describing tetracritical Ising model. After all, we know the IR theory is given by the tetracritical Ising model, and TDLs in UV and IR are matched as
    \begin{equation}
    \begin{array}{ccccccc}
    \text{UV}:&\mcal L_{1,1}&\mcal L_{2,1}&\mcal L_{3,1}&\mcal L_{4,1}&\mcal L_{5,1}\\
    &\downarrow&\downarrow&\downarrow&\downarrow&\downarrow\\
    \text{IR}:&\mcal L_{1,1}&\mcal L_{4,4}&\mcal L_{4,3}&\mcal L_{4,2}&\mcal L_{4,1}
    \end{array},\label{matchingpentaTDLs}
    \end{equation}
    while the other five TDLs
    \[ \{\mcal L_{2,1},\mcal L_{2,2},\mcal L_{3,1},\mcal L_{3,2},\mcal L_{3,3}\} \]
    should be emergent. This result can be used in the opposite direction; as we mentioned in subsection \ref{general}, given the knowledge of IR theory, this result gives a conjecture on classification of MTCs. Namely, there would be no MTC satisfying three conditions simultaneously; (i) which has rank $5<r<10$; (ii) which accommodates central charge smaller than $c_\text{UV}=\frac67$; (iii) which enlarges the surviving MTC $\mcal C$.
    \item $m=7$: Hexacritical Ising model\\
    In this case, the Kac indices are given by
    \begin{align*}
    E_{8,7}=\{&(1,1),(2,1),(2,2),(3,1),(3,2),(3,3),(4,1),(4,2),(4,3),(4,4),\\
    &(5,1),(5,2),(5,3),(5,4),(5,5),(6,1),(6,2),(6,3),(6,4),(6,5),(6,6)\}.
    \end{align*}
    Among these, only six TDLs with
    \[ \widetilde E_{8,7}=\{(1,1),(2,1),(3,1),(4,1),(5,1),(6,1)\} \]
    survive the relevant $\phi_{1,3}$-deformation. In this case, $(r,s)=(6,1)$ gives $\lambda=1$ for all Kac indices in $\widetilde E_{8,7}$. In other words, the symmetric centralizer is non-trivial:
    \begin{equation}
    \text{Obj}Z_2(\mcal C)=\{\mcal L_{1,1},\mcal L_{6,1}\}.\label{hexa'Z2}
    \end{equation}
    Therefore, we claim symmetry should enhance.\footnote{Since the symmetric centralizer is nontrivial, the surviving BFC is not modular. Let us see why studying the braiding is important. The fusion relations of the BFC $\mcal C$ is given by
    \begin{table}[H]
    \begin{center}
    \begin{tabular}{c|c|c|c|c|c|c}
	&$\mcal L_{1,1}$&$\mcal L_{2,1}$&$\mcal L_{3,1}$&$\mcal L_{4,1}$&$\mcal L_{5,1}$&$\mcal L_{6,1}$\\\hline
	$\mcal L_{1,1}$&$\mcal L_{1,1}$&$\mcal L_{2,1}$&$\mcal L_{3,1}$&$\mcal L_{4,1}$&$\mcal L_{5,1}$&$\mcal L_{6,1}$\\\hline
	$\mcal L_{2,1}$&&$\mcal L_{1,1}+\mcal L_{3,1}$&$\mcal L_{2,1}+\mcal L_{4,1}$&$\mcal L_{3,1}+\mcal L_{5,1}$&$\mcal L_{4,1}+\mcal L_{6,1}$&$\mcal L_{5,1}$\\\hline
	$\mcal L_{3,1}$&&&$\mcal L_{1,1}+\mcal L_{3,1}+\mcal L_{5,1}$&$\mcal L_{2,1}+\mcal L_{4,1}+\mcal L_{6,1}$&$\mcal L_{3,1}+\mcal L_{5,1}$&$\mcal L_{4,1}$\\\hline
	$\mcal L_{4,1}$&&&&$\mcal L_{1,1}+\mcal L_{3,1}+\mcal L_{5,1}$&$\mcal L_{2,1}+\mcal L_{4,1}$&$\mcal L_{3,1}$\\\hline
	$\mcal L_{5,1}$&&&&&$\mcal L_{1,1}+\mcal L_{3,1}$&$\mcal L_{2,1}$\\\hline
	$\mcal L_{6,1}$&&&&&&$\mcal L_{1,1}$ 
    \end{tabular}.
    \end{center}
    \end{table}
    This coincides with the one in \cite{rank6} with $SU(2)_5$ realization, which is modular, under identifications
    \[ \phi_{1,1}^\text{here}=1^\text{there},\quad\phi_{2,1}^\text{here}=\phi_4^\text{there},\quad\phi_{3,1}^\text{here}=\phi_3^\text{there},\quad\phi_{4,1}^\text{here}=\phi_5^\text{there},\quad\phi_{5,1}^\text{here}=\phi_2^\text{there},\quad\phi_{6,1}^\text{here}=\phi_1^\text{there}. \]
    So even if two fusion categories share the same fusion relations, one can be modular and another can be non-modular.}
    Again, we cannot explicitly check that a consistent MTC enlarging the surviving rank six BFC $\mcal C$ has rank 15 due to the lack of classification of higher rank MTCs. However, the very fact that $\mcal C$ should be enlarged is consistent with the known RG flow to the pentacritical Ising model. Matching the quantum dimensions, we can identify TDLs in UV and IR theories as follows:
    \begin{equation}
    \begin{array}{ccccccc}
    \text{UV}:&\mcal L_{1,1}&\mcal L_{2,1}&\mcal L_{3,1}&\mcal L_{4,1}&\mcal L_{5,1}&\mcal L_{6,1}\\
    &\downarrow&\downarrow&\downarrow&\downarrow&\downarrow&\downarrow\\
    \text{IR}:&\mcal L_{1,1}&\mcal L_{5,5}&\mcal L_{5,4}&\mcal L_{5,3}&\mcal L_{5,2}&\mcal L_{5,1}
    \end{array}.\label{matchinghexaTDLs}
    \end{equation}
    Counting ranks, we learn the other nine TDLs
    \[\{\mcal L_{2,1},\mcal L_{2,2},\mcal L_{3,1},\mcal L_{3,2},\mcal L_{3,3},\mcal L_{4,1},\mcal L_{4,2},\mcal L_{4,3},\mcal L_{4,4}\} \]
    in IR are emergent. This result also gives a conjecture on classification of MTCs; there is no MTC with rank $r$ in the range $6<r<15$ accommodating central charge $c$ smaller than $c_\text{UV}=\frac{25}{28}$ and enlarging $\mcal C$.
\end{itemize}

\section{Conclusion}
We proposed when and why symmetry enhancements happen in massless RG flows to two-dimensional RCFTs; when the surviving BFC is non-modular or inconsistent with the $c$-theorem, the BFC should be enlarged to an MTC which simultaneously satisfies the two requirements.\footnote{It may be interesting to consider massless RG flows as ``partial functors'' on MTCs by discarding broken objects in the domain. It introduces a natural order in MTCs.} Indeed, these two requirements correctly explain all symmetry enhancements we studied. In particular, we focused on massless RG flows from unitary minimal models in order to check our proposal. However, since these two conditions $-$ modularity and the $c$-theorem $-$ are well established in two-dimensional RCFTs, we believe our proposal works in more general cases. Then our proposal starts to either `predict' symmetry enhancements, or give conjectures on classification of MTCs. Let us list some interesting problems.
\begin{itemize}
    \item $M(2M+2,2M+1)\pm\phi_{r,s}$ with odd $r,s$\\
    We have studied $\phi_{1,3}$-deformation in detail, and in particular saw $\mbb Z_2$ symmetry is preserved by the operator. More generally, one can show $\phi_{r,s}$ with odd $r,s$ preserves the $\mbb Z_2$ symmetry generated by $\mcal L_{m-1,1}$.\footnote{To prove this, just calculate the action $\hat{\mcal L}_{m-1,1}|\phi_{r,s}\ra$:
    \begin{align*}
        \hat{\mcal L}_{m-1,1}|\phi_{r,s}\ra&=\frac{S_{(m-1,1),(r,s)}}{S_{(1,1),(r,s)}}|\phi_{r,s}\ra\\
        &=(-1)^{m(r+s)}\frac{\sin\frac{\pi(m-1)r}m}{\sin\frac{\pi r}m}|\phi_{r,s}\ra=-(-1)^{r+m(r+s)}|\phi_{r,s}\ra.
    \end{align*}
    So primaries with odd $r,s$ commute with $\mcal L_{m-1,1}$.}
    Furthermore, when $m=2M+1$, the preserved TDL $\mcal L_{m-1,1}$ can give a nontrivial element of the symmetric centralizer. Although the $\phi_{3,3}$-perturbation of tetracritical Ising model ($m=5$) seem to trigger massive RG flows regardless of the sign of $\lambda_{3,3}$ \cite{Ma}, more general $\phi_{r,s}$-deformation with odd $r,s$ may trigger massless RG flows for larger $m$. In that case, our proposal claims its symmetry should enhance.
    \item Three-state Potts model\\
    Our proposal would work not just in diagonal minimal models, but also in more general RCFTs. One slightly general example is the three-state Potts model, which is obtained from the tetracritical Ising model by gauging the $\mbb Z_2$ symmetry. One can show the symmetric centralizers are always trivial under any perturbations of the theory. So if the perturbations trigger massless RG flows to RCFTs, symmetries do not have to enhance unless central charges are inconsistent with the $c$-theorem.
    \item Wess-Zumino-Witten models\\
    Another class of RCFTs is Wess-Zumino-Witten (WZW) models. Unlike minimal models \cite{CW}, some WZW models have 't Hooft anomalies \cite{GWi,NY,LS,KZ}. So one can constrain IR theories more severely incorporating 't Hooft anomaly matching conditions \cite{tH,FO,YHO,TS}. If anomalies forbid massive phases, one can get predictions of symmetry enhancements with weaker assumptions (just rationality).
    \item Irrational CFT to RCFT\\
    So far, we assumed not only IR CFTs but also UV CFTs to be rational. However, we can relax the latter assumption; our proposal should also apply to massless RG flows from a UV irrational CFT to an IR RCFT because we know the latter is described by an MTC. To tackle this problem, the approach initiated in \cite{CY} may be helpful.
    \item Fermionic CFTs\\
    There are also RCFTs with non-integer spins. Recently, fermionic counterparts of minimal models were constructed \cite{HNT,K}. It would be interesting to generalize our proposal to accommodate these examples.
\end{itemize}
It is also desirable to extend our idea to other dimensions. Although it is unclear what kind of categories (or more general notion) are appropriate in other dimensions, we have generalization of the $c$-theorem in other dimensions:
\begin{itemize}
    \item $F$-theorem \cite{JKPS,CDFKS} in three dimensions
    \item $a$-theorem \cite{C88,KS,K11} in four dimensions
    \item $a$-theorem \cite{CDY,CDI} in six dimensions
\end{itemize}
So optimistically, combinations of the generalized $c$-theorems and appropriate mathematical notions would also explain symmetry enhancements in other dimensions. As conformal symmetry plays the crucial role in our proposal, a natural strategy to tackle these generalizations is to impose larger spacetime symmetries, such as supersymmetry, as is usually done in four dimensions \cite{RSZ,SZ,HPS}. In addition, it would be instructive first to restrict oneself to systems without continuous symmetries. We would like to leave these interesting problems for the future.

\section*{Acknowledgment}
We thank Yuji Tachikawa for helpful discussions and suggestions from the early stage of the project. We also thank Chi-Ming Chang for helpful comments on the draft.

\appendix
\setcounter{section}{0}
\renewcommand{\thesection}{\Alph{section}}
\setcounter{equation}{0}
\renewcommand{\theequation}{\Alph{section}.\arabic{equation}}

\section{Review of minimal models}\label{notation}
For a natural number $m\ge3$, one has a unitary minimal model $M(m+1,m)$\footnote{We basically follow notions of \cite{FMS}.} with central charge
\[ c=1-\frac6{m(m+1)}. \]
The theory has conformal primaries $\phi_{r,s}$ labeled by Kac indices $(r,s)$ in the range\footnote{Another way to define the Kac table is to relax the third condition, and impose the equivalence relation
\begin{equation}
    (r,s)\sim(m-r,m+1-s).\label{Kaceqrel}
\end{equation}}
\begin{equation}
    E_{m+1,m}:=\{(r,s)|1\le r\le m-1\&1\le s\le m\&ms<(m+1)r\}.\label{Kac}
\end{equation}
We stick to this notation with one exception; we label the deformation operator and coupling with $(r,s)=(1,3)$ instead of $(r,s)=(m-1,m-2)$. Note that although $s=m$ seems to belong to the set just from the second defining condition, the third condition $ms<(m+1)r\le m^2-1$ excludes $s=m$. Hence, the number of primaries in the theory is
\begin{equation}
    \frac12m(m-1).\label{numprim}
\end{equation}
It is known that there is a one-to-one correspondence between the primaries and topological defect lines (TDLs), called Verlinde lines \cite{V}, in minimal models (or more generally in diagonal RCFTs). We also label the TDLs with Kac indices as $\mcal L_{r,s}$. The modular $S$-matrix of the theory is given by\footnote{One can check the modular $S$-matrix is invariant under $(r,s)\leftrightarrow(m-r,m+1-s)$.}
\begin{equation}
S_{(r,s),(r',s')}=(-1)^{(r+s)(r'+s')}\sqrt{\frac8{m(m+1)}}\sin\left(\frac{\pi rr'}m\right)\sin\left(\frac{\pi ss'}{m+1}\right).\label{minS}
\end{equation}
Employing the modular $S$-matrix, one can show $\mcal L_{m-1,1}$ always generates $\mbb Z_2$ symmetry of the theory as we will see in Appendix \ref{MTC}. In fact, the $\mbb Z_2$ symmetry is inherited from minimal models at $m\to+\infty$ down to $m=3$.

The action of TDLs on primaries can be computed using the modular $S$-matrix. The action is given by
\begin{equation}
\hat{\mcal L}_i|\phi_j\ra=\frac{S_{ij}}{S_{0j}}|\phi_j\ra.\label{actionL}
\end{equation}
If a TDL $\mcal L_i$ acts on an operator $\phi_j$ as on the identity operator
\begin{equation}
    \frac{S_{ij}}{S_{0j}}=\frac{S_{i0}}{S_{00}},\label{commute}
\end{equation}
the TDL is said to commute with the operator. If a TDL commute with a relevant operator, the TDL does not `feel' an insertion of the operator. So the TDL survives all along the RG flow triggered by the operator \cite{TDL}. In the main part of this paper, we study relevant deformation with $\phi_{1,3}$. Therefore, surviving TDLs are labeled by Kac indices
\begin{equation}
\widetilde E_{m+1,m}:=\left\{(r,s)\in E_{m+1,m}\Big|\frac{\sin\frac{3\pi s}{m+1}}{\sin\frac{3\pi}{m+1}}=\frac{\sin\frac{\pi s}{m+1}}{\sin\frac\pi{m+1}}\right\}.\label{survKac}
\end{equation}
Note that the identity $(r,s)=(1,1)$ trivially survives. Interestingly, one immediately realizes that not only the identity but also any TDLs with Kac indices $s=1$ automatically survive the $\phi_{1,3}$-deformation. In fact, one can prove the only solutions are $s=1$ (see the Appendix \ref{proof}). So we get an explicit expression:
\begin{equation}
\widetilde E_{m+1,m}=\{(r,1)|r=1,2,\dots,m-1\}.\label{survKac'}
\end{equation}
Especially, the number of surviving TDLs is given by $(m-1)$.

Using these formulae, we can formally state when symmetry should enhance: \textit{$\exists(r_*,s_*)\in\widetilde E_{m+1,m}\backslash(1,1)$ s.t. $\forall(r,s)\in\widetilde E_{m+1,m}$, $c_{(r_*,s_*),(r,s)}\circ c_{(r,s),(r_*,s_*)}=id_{\mcal L_{(r,s)\otimes(r_*,s_*)}}$.} By studying the braiding in detail later, we simplify this condition; symmetry should enhance when $m>3$ is odd, but for even $m$, the symmetric centralizer is trivial, and one has to compare central charges in UV and in IR.

\section{Braided Fusion Category, Modular Tensor Category, and braiding matrix}\label{MTC}
In this appendix, we give a minimal background to understand our results. Since we focus on category symmetries in two dimensions, we give the definitions in terms of topological defect lines (TDLs). For more details, see e.g. \cite{DGNO,Mu} for mathematics- and \cite{FFRS,BT,TDL} for physics-oriented literature. Using the knowledge, we also compute braiding matrices, the central tool we use in this paper.

\subsection{Definitions}
A braided fusion category (BFC) is a fusion category $\mcal C$ equipped with a braiding $c$ which is a natural isomorphism on two objects $\mcal L_i,\mcal L_j\in\text{Obj}(\mcal C)$:
\begin{equation}
    c_{i,j}:\mcal L_i\otimes\mcal L_j\simeq\mcal L_j\otimes\mcal L_i.\label{braiding}
\end{equation}
Performing it twice
\[ c_{j,i}\circ c_{i,j}:\mcal L_i\otimes\mcal L_j\to \mcal L_i\otimes\mcal L_j, \]
one comes back with TDLs $\mcal L_i$ and $\mcal L_j$ braided, hence the name. Note that the full braiding is in general described by a matrix.\footnote{When $\mcal L_i\otimes\mcal L_j$ is simple, it becomes
\[ c_{j,i}\circ c_{i,j}=\lambda_{i,j}\cdot id_{\mcal L_i\otimes\mcal L_j} \]
for a scalar $\lambda_{i,j}$.}
If the isomorphism is trivial
\[ c_{j,i}\circ c_{i,j}=id_{\mcal L_i\otimes\mcal L_j}, \]
$\mcal L_i$ and $\mcal L_j$ are said to commute.
If $\mcal L_i$ commutes with all objects, $\mcal L_i$ is called transparent. Collection of transparent objects defines the symmetric centralizer
\begin{equation}
    \text{Obj}Z_2(\mcal C):=\{\mcal L_i\in\text{Obj}(\mcal C)|\forall\mcal L_j\in\text{Obj}(\mcal C),c_{j,i}\circ c_{i,j}=id_{\mcal L_i\otimes\mcal L_j}\}.\label{Z_2}
\end{equation}
A BFC with trivial symmetric centralizer is called non-degenerate. The name can be understood after we introduce pivotal structure.

A pivotal structure of a fusion category $\mcal C$ is an isomorphism of tensor functors
\[ \gamma:id\to**, \]
where $*$ is the dual of the fusion category.\footnote{Physically, the dual corresponds to reversing the orientation of TDLs.} It is conjectured that all fusion categories admit pivotal structures \cite{ENO}. A fusion category equipped with a pivotal structure is called a pivotal category.

Given a pivotal structure, one can define left and right traces of $s$, $\tr_{\mcal L}^L(s)$ and $\tr_{\mcal L}^R(s)$, where $\mcal L\in\text{Obj}(\mcal C)$ and $s\in\text{End}\mcal L$. If the two traces coincide for all $\mcal L\in\text{Obj}(\mcal C)$, i.e., $\tr_{\mcal L}^L=\tr_{\mcal L}^R$, the pivotal category is called spherical. So on a spherical pivotal category, one can simply talk about a trace $\tr=\tr^L=\tr^R$. Using the trace, we define quantum dimensions\footnote{In computing fusion relations, it is useful to know quantum dimensions obey the same relation
\begin{equation}
d_id_j=\sum_k{N_{ij}}^kd_k\label{qdproduct}
\end{equation}
as shown, for example, in Proposition 4.9 of \cite{qdproduct}. Using this relation, one can show invertible TDLs should have quantum dimension one; an invertible TDL $\mcal L$ has its inverse $\mcal L^{-1}$, and they fuse to the trivial TDL. So evaluation of the quantum dimensions give
\[ d_{\mcal L}d_{\mcal L^{-1}}=d_{id}=1. \]
Therefore, taking $d_i\ge1$ into account, we need $d_{\mcal L}=1$.} of objects as
\begin{equation}
    d_i:=\tr_{\mcal L_i}(\gamma).\label{qd}
\end{equation}
If a braided fusion category is spherical, it is called pre-modular. The reason behind the name is this; on a pre-modular BFC $\mcal C$, one can define an $S$-matrix\footnote{One may wonder how modular $T$-matrices are defined. Mathematically, they are defined by a ribbon structure
\[ \Theta:id_{\mcal C}\to id_{\mcal C}, \]
or
\[ \Theta_{\mcal L}=(\tr_{\mcal L}\otimes id_{\mcal L})(c_{\mcal L,\mcal L}), \]
and the $T$-matrix is given by
\[ \widetilde T_{i,j}=\delta_{i,j}\Theta_{\mcal L_i}. \]
In physics, however, we usually include a phase depending on the central charge:
\[ T_{i,j}=e^{-2\pi ic/24}\widetilde T_{i,j}. \]}
\begin{equation}
    \mcal L_i,\mcal L_j\in\text{Obj}(\mcal C),\quad\widetilde S(\mcal L_i,\mcal L_j):=\tr_{\mcal L_i\otimes\mcal L_j}(c_{j,i}\circ c_{i,j}).\label{Smatrix}
\end{equation}
In physics, we usually use normalized $S$-matrix $\widetilde S/D$ where $D^2:=\sum_id_i^2$ is the total quantum dimension (squared).\footnote{In math literature, it is called the (global) dimension of $\mcal C$:
\[ \dim(\mcal C):=D^2. \]} In general, the $S$-matrix is non-invertible. However, if $\mcal C$ is non-degenerate, i.e., its symmetric centralizer is trivial, the $S$-matrix is invertible. This finishes the list of definitions; a BFC is called modular if it is both pre-modular and non-degenerate. Modular categories are also called modular tensor categories (MTCs).

In this paper, our starting point is a UV RCFT with an MTC. We consider its perturbation which preserves a braided fusion subcategory $\mcal C$. Along the RG flow, fusion relations, quantum dimensions, and braidings are preserved. Furthermore, since $\mcal C$ is a subcategory of the UV MTC, it is automatically spherical. Therefore, whether the BFC $\mcal C$ is an MTC or not is determined by whether its symmetric centralizer $Z_2(\mcal C)$ is trivial or not. To judge modularity of $\mcal C$, we study its braiding matrix $\lambda$.

\subsection{Braiding matrix}\label{braid}
Let us consider a pre-modular BFC $\mcal C$. If all $\mcal L_i,\mcal L_j,\mcal L_i\otimes\mcal L_j$ are simple, the braiding matrix is given by (see for example \cite{braiding})
\begin{equation}
\lambda_{i,j}=\frac{\widetilde S_{i,j}}{d_id_j},\label{lambda}
\end{equation}
where $\widetilde S$ is the unnormalized modular $S$-matrix $\widetilde S_{i,j}:=DS_{i,j}$ with $S_{i,j}$ the usual normalized modular $S$-matrix. Now, let us apply the formulation to our problem, minimal models.

With the modular $S$-matrix (\ref{minS}) at hand, it is not hard to compute the braiding matrices. First, the quantum dimensions are given by
\begin{equation}
    d_i=\frac{S_{0i}}{S_{00}},\label{qd'}
\end{equation}
so
\begin{equation}
d_{(r,s)}=\frac{\sin\frac{\pi r}m\sin\frac{\pi s}{m+1}}{\sin\frac\pi m\sin\frac\pi{m+1}}.\label{minqd}
\end{equation}
in the minimal model $M(m+1,m)$.\footnote{It is worth mentioning that $(r,s)=(m-1,1)$ always has quantum dimension one. Using the fact, one immediately learns the TDL $\mcal L_{m-1,1}$ always generates $\mbb Z_2$ symmetry of $M(m+1,m)$. To show this, let us study the fusion of $\mcal L_{m-1,1}$ with itself. Because of the symmetry of the fusion coefficient, we know ${N_{id,(m-1,1)}}^{(m-1,1)}=1={N_{(m-1,1),(m-1,1)}}^{id}$, i.e., $\mcal L_{m-1,1}\mcal L_{m-1,1}=\mcal L_{id}+\cdots.$ Now, evaluating quantum dimensions of both hand sides, we already get $1=1+\cdots$, so no other TDLs can appear on the RHS. Hence, $\mcal L_{m-1,1}$ generates $\mbb Z_2$ symmetry.

We would also like to mention the quantum dimension of $\mcal L_{2,1}$, which always survive the $\phi_{1,3}$-deformation. Its quantum dimension is
\[ d_{2,1}=\frac{\sin\frac{2\pi}m}{\sin\frac\pi m}=2\cos\frac\pi m. \]
This is one iff $m=3$. Thus, when $m>3$, $d_{2,1}$ is not one, and in particular not a non-negative integer. So we can employ the criterion of \cite{TDL}, and immediately conclude there exist degenerated vacua in massive RG flows. Since $\mcal L_{2,1}$ is non-invertible (for $m>3$), the degeneracy is a consequence of spontaneous non-invertible symmetry breaking. We will use this fact in Appendix \ref{massive}.} In addition, we can also compute the braiding matrix explicitly:
\[ \lambda_{(r,s),(r',s')}=(-1)^{(r+s)(r'+s')}\frac{\sin\frac{\pi rr'}m\sin\frac{\pi ss'}{m+1}\sin\frac\pi m\sin\frac\pi{m+1}}{\sin\frac{\pi r} m\sin\frac{\pi s}{m+1}\sin\frac{\pi r'}m\sin\frac{\pi s'}{m+1}}. \]
(Remember that this formula is valid if two simple lines fuse to a simple line.) As a check, one can easily see the matrix is symmetric. In addition, when $(r,s)=(1,1)$ or $(r',s')=(1,1)$, the $\lambda$'s are all 1. The identity $(1,1)$ gives the trivial element of the symmetric centralizer.

We study this braiding matrix in more detail below. In particular, we try to find transparent line(s) to check modularity of surviving BFCs. One notices at first that the formula is applicable when one of the two fusing lines is invertible, i.e., $(r,s)=(m-1,1)$ or $(r',s')=(m-1,1)$. This is because invertibility guarantees the resulting line is also simple. The braiding matrix in this case reduces to
\begin{equation}
    \lambda_{(r,s),(m-1,1)}=-(-1)^{r+m(r+s)}.\label{lambdaZ_2}
\end{equation}
We study the cases for odd and even $m$ in turn.

\subsubsection{Odd $m$}
For odd $m=2M+1$ ($M=2,3,\dots$), one can show a TDL with $(r_*,s_*)=(m-1,1)$ always gives a nontrivial symmetric centralizer. To prove this claim, just insert $m=2M+1$ in (\ref{lambdaZ_2}):
\begin{equation}
\lambda_{(r,s),(m-1,1)}=-(-1)^{r+(r+s)}=1,\label{lambda=1odd}
\end{equation}
where in the last equality, we used the lemma, i.e., $s=1$ for surviving TDLs. Therefore, our proposal claims symmetry should enhance when $m$ (larger than three) is odd.

\subsubsection{Even $m$}
In this case $m=2M$ ($M=2,3,\dots$), we can show the symmetric centralizer is always trivial. The idea is to constrain possible values of $(r_*,s_*)$. Recall the definition of symmetric centralizer; if $(r_*,s_*)$ gives a nontrivial element, it should have $c_{(r_*,s_*),(r,s)}\circ c_{(r,s),(r_*,s_*)}=id_{\mcal L_{(r,s)}\otimes\mcal L_{(r_*,s_*)}}$ for all $(r,s)$ in $\widetilde E_{m+1,m}$. Therefore we consider each $(r,s)\in\widetilde E_{m+1,m}$ one by one. Firstly, let us consider $(r,s)=(1,1)$. As we saw above, this always gives $\lambda=1$, and does not impose any constraint on $(r_*,s_*)$. Next, let us study the case $(r,s)=(m-1,1)$. Since $m\ge3$, this is always an element of $\widetilde E_{m+1,m}$. Furthermore, since the line is invertible, we can still use (\ref{lambdaZ_2}). The formula in this case reduces to
\begin{equation}
    1\stackrel!=\lambda_{(r_*,s_*),(m-1,1)}=-(-1)^{r_*},\label{lambda=1even1}
\end{equation}
Thus transparent lines should have odd $r_*$. The second equality tells us that the $\mbb Z_2$ line $\mcal L_{2M-1,1}$ itself is not transparent because we always have a surviving line $(r,1)$ with even $r$, say, $r=2$.

Since the minimal model has only two invertible lines, we can no longer rely on the explicit braiding matrix to impose further constraints on $(r_*,s_*)$. Here, we instead employ a more detailed technique of BFC. In particular, we use the Lemma E.13 of \cite{K06}; a TDL $(r_*,s_*)$ is transparent in $\mcal C$ iff for all $\mcal L\in\text{Obj}(\mcal C)$, $\mcal L$ commutes with $\phi_{r_*,s_*}$. Again, we try to constrain possible $(r_*,s_*)$ by considering each $(r,s)\in\widetilde E_{2M+1,2M}$ one by one. From the result above, we only have to consider $(r_*,s_*)=(t,1)$ with odd $t=2T-1$. Since $\mcal L_{2M-1,1}$ is not transparent, possible values of $T$ is $T=1,\dots,M-1$. Using the action (\ref{actionL}), we can express the commutativity condition as $\frac{S_{(r,1),(2T-1,1)}}{S_{(1,1),(2T-1,1)}}=\frac{S_{(r,1),(1,1)}}{S_{(1,1),(1,1)}}$, or
\[ \frac{\sin\frac{\pi r(2T-1)}{2M}}{\sin\frac{\pi(2T-1)}{2M}}=\frac{\sin\frac{\pi r}{2M}}{\sin\frac\pi{2M}}. \]
The condition is trivially satisfied for $(r,1)=(1,1)$, and does not impose any constraint on $T$. Next, let us consider $(r,1)=(2,1)$. Using the double-angle formula, we get a reduced condition
\[ \cos\frac{\pi(2T-1)}{2M}=\cos\frac\pi{2M}. \]
This can be solved easily:
\[ \frac{\pi(2T-1)}{2M}=\pm\frac\pi{2M}+2\pi n\quad(n\in\mbb Z). \]
In the range $T=1,\dots,M-1$, the only solution is $T=1$,\footnote{For $T$ to be in the range, we first of all need $n=0$. Then plus sign gives $T=1$, while negative sign gives $T=0$.} or $(r_*,s_*)=(1,1)$. In summary, we proved the symmetric centralizer of surviving BFC is always trivial for even $m$.

\section{Proof of the lemma}\label{proof}
In this appendix, we prove the Kac indices of the surviving TDLs are given by $(r,1)$:
\begin{equation}
\widetilde E_{m+1,m}=\left\{(r,s)\in E_{m+1,m}\Big|\frac{\sin\frac{3\pi s}{m+1}}{\sin\frac{3\pi}{m+1}}=\frac{\sin\frac{\pi s}{m+1}}{\sin\frac\pi{m+1}}\right\}=\{(r,1)|r=1,2,\dots,m-1\}.\label{lemma}
\end{equation}
Let us apply the triple-angle and double-angle formula on the defining condition of $\widetilde E_{m+1,m}$:
\begin{align*}
\frac{\sin\frac{3\pi s}{m+1}}{\sin\frac{3\pi}{m+1}}&=\frac{\sin\frac{\pi s}{m+1}\left(3-4\sin^2\frac{\pi s}{m+1}\right)}{\sin\frac\pi{m+1}\left(3-4\sin^2\frac\pi{m+1}\right)}\\
&=\frac{\sin\frac{\pi s}{m+1}}{\sin\frac\pi{m+1}}\frac{1+2\cos\frac{2\pi s}{m+1}}{1+2\cos\frac{2\pi}{m+1}}.
\end{align*}
Thus, the condition reduces to
\[ 0=\frac{\sin\frac{\pi s}{m+1}}{\sin\frac\pi{m+1}}\left(\frac{1+2\cos\frac{2\pi s}{m+1}}{1+2\cos\frac{2\pi}{m+1}}-1\right). \]
Since the overall factor is non-zero in the range of $s$, the defining condition reduces to a much simpler form:
\begin{equation}
\cos\frac{2\pi s}{m+1}=\cos\frac{2\pi}{m+1}.\label{surv}
\end{equation}
This can be easily solved:
\[ \frac{2\pi s}{m+1}=\pm\frac{2\pi}{m+1}+2\pi n\quad(n\in\mbb Z). \]
In the range $s\in\{1,2,\dots,m-1\}$, the only solution is $s=1$,\footnote{A solution with the plus sign requires $n=0$ because otherwise $s$ is out of the range. So this choice gives $s=1$. If one tries to find a solution with the minus sign, one notices that any choice of $n$ takes $s$ out of the range. Hence there is no solution with the minus sign.} showing the lemma.

\section{Massive RG flows}\label{massive}
To study massive (and massless) RG flows, we employ Cardy's method \cite{Cardy}. So let us first recall the definition of Cardy states. For each primary field, there exists conformal boundary state (also called Cardy state \cite{Cstate})
\begin{equation}
|i\ra:=\sum_j\frac{S_{ij}}{\sqrt{S_{0j}}}|j\ra\ra,\label{Cardystate}
\end{equation}
where $|j\ra\ra$ is the Ishibashi state \cite{I88,OI} corresponding to the irrep $j$. One property we will use below is the linear independence of Cardy states:
\[ \la i|j\ra\sim\delta_{i,j}. \]
On Cardy states, TDLs can act. To get the transformation law, we only have to know how Ishibashi states transform; they behave as the corresponding primary operator:
\begin{equation}
\hat{\mcal L_i}|j\ra\ra=\frac{S_{ij}}{S_{0j}}|j\ra\ra.\label{actionLonIshibashi}
\end{equation}
Using the action, one finds Cardy states transform consistently with fusion relations \cite{GWa}:
\begin{equation}
\hat{\mcal L_i}|j\ra=|i\times j\ra=\sum_k{N_{ij}}^k|k\ra.\label{actionLonCardy}
\end{equation}
Since we are interested in ground state degeneracy in the case of massive RG flows, we would like to ask whether a ground state is invariant under all surviving TDLs. If it is invariant under all surviving TDLs, then there would be no ground state degeneracy. On the other hand, if there exists a TDL under which the ground state is charged, then it signals spontaneous breaking of the symmetry generated by the TDL \cite{CO}, and there would be ground state degeneracies. In particular, if a ground state $|a\ra$ is charged under a TDL $\mcal L$, then $|a\ra$ and $\hat{\mcal L}|a\ra\neq|a\ra$ have degenerated energy because $\mcal L$ commutes with Hamiltonian by definition. Recalling (\ref{actionLonCardy}), we can diagnose whether a ground state $|a\ra$ is degenerated or not by studying fusion rules of the primary $\phi_a$.

Armed with this observation, let us study when a ground state is non-degenerate. From the discussion above, the ground state should be invariant under the action of all surviving TDLs:
\begin{equation}
\forall i\in\widetilde E_{m+1,m},\quad\hat{\mcal L}_i|j\ra=|j\ra.\label{invCardy}
\end{equation}
Evaluating quantum dimensions of
\[ \mcal L_i\mcal L_j=\mcal L_j, \]
we find a necessary condition:
\begin{equation}
\forall i\in\widetilde E_{m+1,m},\quad d_i=1.\label{necessary}
\end{equation}
However, as we saw above, $\mcal L_{2,1}$ always has quantum dimension $d_{2,1}\neq1$ for $m>3$. Therefore, when $m>3$, there is no Cardy state invariant under all surviving TDLs.\footnote{This is not the case for $m=3$, i.e., the critical Ising model. Indeed, the spin operator $\sigma$ `absorbs' primary operators corresponding to surviving TDLs:
\[ id\times\sigma=\sigma,\quad\vep\times\sigma=\sigma. \]
So $|\sigma\ra=|(2,2)\ra$ gives an invariant Cardy state.} (We also comment that there may be Cardy states invariant under a subset of TDLs with quantum dimensions one.) Therefore, for $m>3$, there exists a TDL under which the ground state $|a\ra$ is charged. Since the theory is topological, this means spontaneous breaking of the TDLs \cite{CO}. This also implies an existence of ground state degeneracy.

Now, let us examine which Cardy states can be ground states in the IR TQFT, the fixed point of our massive RG flow from a minimal model perturbed by $\phi_{1,3}$. It is known \cite{Cardy} that depending on the sign of the relevant coupling $\lambda_{1,3}$, the Cardy states are given by either $s=1$ or $s\simeq\frac m2$. More precisely, the ground state(s) minimize the product
\[ \left(1+2\cos\frac{2\pi s}{m+1}\right)\lambda_{1,3}, \]
where $\lambda_{1,3}$ is the relevant coupling entering the Hamiltonian as $H\ni\lambda_{1,3}\int\phi_{1,3}$. Let us study negative and positive $\lambda_{1,3}$ in turn.

\begin{itemize}
    \item $\lambda_{1,3}<0$:\\
    In this case, the ground state(s) are given by maximizing the first factor. In our convention of Kac indices, this is achived by $s=1$. Then it is clear that $(m-1)$ Kac indices $(r,1)$ with $r=1,2,\dots,m-1$ give candidate ground states:
    \[ E_C^{\lambda_{1,3}<0}=\{(r,1)|r=1,2,\dots,m-1\}. \]
    Recalling the one-to-one correspondences among Verlinde lines, primaries, and Cardy states, it is natural to think this sign (positive in the Lagrangian language) gives massless RG flows (we will see the other sign does not accommodate $(1,1)$). The sign is consistent with \cite{Ma}.
    \item $\lambda_{1,3}>0$:\\
    In this case, we have to minimize the first factor. This is done by $s$ such that $\frac{2\pi s}{m+1}$ is the closest to $\pi$ mod $2\pi$. To find which $s$ gives the lowest energy, let us perform case analysis.
    \begin{itemize}
        \item Odd $m$:\\
        Let us denote $m=2M+1$ with $M=1,2,\dots$. The condition reduces to
        \[ s\simeq(M+1)+2(M+1)n\quad(n\in\mbb Z). \]
        For $s$ to be in the range $1,2,\dots,m-1=2M$, we need $n=0$, and $s=M+1$ minimizes the first factor, hence the product. The candidate Cardy states are given by $M$ Kac indices\footnote{Note that this is different from $(m-1)=2M$ suggested in the literature. Previous results support $2M$-fold degeneracy, however, Cardy's method suggests $M$-fold degeneracy. We checked the first factor explicitly for $M=1,2,3$, and we always got $M$. So at present we are not sure which is correct, whether $M$ is the true ground state degeneracy, or we are applying Cardy's method beyond its validity.\label{rmk}}
        \[ E_C^{\lambda_{1,3}>0\&m=2M+1}=\{(r,M+1)|r=M+1,M+2,\dots,2M\}. \]
        \item Even $m$:\\
        We write $m=2M+2$ with $M=1,2,\dots$. The condition reduces to
        \[ s\simeq M+1+\frac12+(2M+3)n\quad(n\in\mbb Z). \]
        Again, for $s$ to be in the range $1,2,\dots,m-1=2M+1$, we need $n=0$. So $s=M+1,M+2$ give the same minimum factor. The candidate Cardy states are given by $(m-1)$ Kac indices
        \begin{align*} E_C^{\lambda_{1,3}>0\&m=2M+2}=\{(r,M+1),(r',M+2)|r=M+1,\dots,2M+1,r'=M+2,\dots,2M+1\}.
        \end{align*}
        Our suggestion is consistent with the known RG flow from tricritical Ising model to the critical Ising model triggered by $-\vep'$ \cite{H}.
    \end{itemize}
\end{itemize}
Since the Cardy states are linearly independent, these explicit candidates of Cardy states give the (maximum) number\footnote{Note that it is not clear a priori that the ground state degeneracy equals the maximum number because some candidate states may not appear. However, we can create all candidate states as follows; we employ the fusion rule
\[ \phi_{2,1}\times\phi_{r,s}=\phi_{r-1,s}+\phi_{r+1,s} \]
where the primaries on the RHS vanish if $(r\pm1)$ are beyond the range of the Kac table. By acting $\mcal L_{2,1}$ repeatedly, we see all states appear. Therefore, the IR TQFT should have all candidate states. This argument suggests that the ground state degeneracy is given by the maximum number of Cardy states.} of ground state degeneracy.

Therefore, we can summarize RG flows from unitary minimal models triggered by the relevant operator $\phi_{1,3}$ as follows:
\begin{itemize}
    \item $\lambda_{1,3}<0$ (or positive coupling in the Lagrangian language):\\
    Massless RG flow to the next minimal model.
    \item $\lambda_{1,3}>0$ (or negative coupling in the Lagrangian language):\\
    Massive RG flow to IR TQFT with ground state degeneracy
    \begin{equation}
        \left\{\begin{split}
            \frac{m-1}2\quad&(m\text{ odd}),\\
            (m-1)\quad&(m\text{ even}).
        \end{split}\right.\label{gsd}
    \end{equation}
\end{itemize}
Our sign proposal is consistent with \cite{Ma}. For even $m$, the ground state degeneracy is also consistent with suggestions in literature, however, for odd $m$, our result gives a half. We are not sure where the mismatch is coming from.

Let us see this general argument in concrete examples.

\subsection{Critical Ising model}
In this case, we have three Cardy states corresponding to three primary fields with Kac indices $(1,1),(2,1),(2,2)$:
\[
\begin{split}
|(1,1)\ra&=\frac1{\sqrt2}\left(|id\ra\ra+|\vep\ra\ra+2^{1/4}|\sigma\ra\ra\right),\\
|(2,1)\ra&=\frac1{\sqrt2}\left(|id\ra\ra+|\vep\ra\ra-2^{1/4}|\sigma\ra\ra\right),\\
|(2,2)\ra&=\frac1{\sqrt2}\left(\sqrt2|id\ra\ra-\sqrt2|\vep\ra\ra\right).
\end{split}
\]
Under the perturbation $\delta H=\lambda_{1,3}\int\phi_{1,3}$, each state has energy \cite{Cardy} (with rescaled coupling)
\begin{align*}
    E_{1,1}&=\frac\pi{48(2\tau_{1,1})^2}+\frac{\lambda_{1,3}}{2\tau_{1,1}},\\
    E_{2,1}&=\frac\pi{48(2\tau_{2,1})^2}+\frac{\lambda_{1,3}}{2\tau_{2,1}},\\
    E_{2,2}&=\frac\pi{48(2\tau_{2,2})^2}-\frac{\lambda_{1,3}}{2\tau_{2,2}}.
\end{align*}
Therefore, for $\lambda_{1,3}<0$ (or positive coupling in Lagrangian language) two Cardy states $|(1,1)\ra,|(2,1)\ra$ give degenerated minimum energy. On the other hand, for $\lambda_{1,3}>0$ (or negative coupling in Lagrangian language) Cardy state $|(2,2)\ra$ gives the minimum energy, hence the ground state. This result is well-known, and supports our counting (\ref{gsd}) $1=\frac{3-1}2$. Note that non-degenerated ground state is possible because the corresponding primary $\phi_{2,2}=\sigma$ is invariant under fusion with primaries corresponding to surviving TDLs:
\[ 1\times\sigma=\sigma,\quad\vep\times\sigma=\sigma. \]
The invariant ground state is realized because the necessary condition (\ref{necessary}) is satisfied. However, this does not happen in general as we saw above.

\subsection{Tricritical Ising model}
As in the previous example, there are six Cardy states corresponding to six primary fields. For example, some of them are given by
\[
\begin{split}
|(1,1)\ra&=\left(\frac{s_1}{\sqrt5}\right)^{1/2}\left[|id\ra\ra+2^{1/4}|\sigma'\ra\ra+2^{1/4}\zeta^{1/2}|\sigma\ra\ra+|\vep''\ra\ra+\zeta^{1/2}|\vep'\ra\ra+\zeta^{1/2}|\vep\ra\ra\right],\\
|(2,1)\ra&=\left(\frac{s_1}{\sqrt5}\right)^{1/2}\left[\sqrt2|id\ra\ra-\sqrt2|\vep''\ra\ra+\sqrt2\zeta^{1/2}|\vep'\ra\ra-\sqrt2\zeta^{1/2}|\vep\ra\ra\right],\\
|(3,1)\ra&=\left(\frac{s_1}{\sqrt5}\right)^{1/2}\left[|id\ra\ra-2^{1/4}|\sigma'\ra\ra-2^{1/4}\zeta^{1/2}|\sigma\ra\ra+|\vep''\ra\ra+\zeta^{1/2}|\vep'\ra\ra+\zeta^{1/2}|\vep\ra\ra\right],\\
\end{split}
\]
and so on. Here, $s_1:=\sin\pi/5$, and $\zeta:=\frac{1+\sqrt5}2$ is the golden ratio.

Let us study which Cardy state gives the minimum energy. Applying the Cardy's method \cite{Cardy} for a perturbation $\delta H=\lambda_{1,3}\int\phi_{1,3}$,
we get energies (with rescaled couplings)
\begin{equation}
\begin{split}
E_{11}&=\frac{7\pi}{240(2\tau_{11})^2}+\zeta\frac{\lambda_{1,3}}{(2\tau_{11})^{6/5}},\\
E_{21}&=\frac{7\pi}{240(2\tau_{21})^2}+\zeta\frac{\lambda_{1,3}}{(2\tau_{21})^{6/5}},\\
E_{22}&=\frac{7\pi}{240(2\tau_{22})^2}-\zeta^{-1}\frac{\lambda_{1,3}}{(2\tau_{22})^{6/5}},\\
E_{31}&=\frac{7\pi}{240(2\tau_{31})^2}+\zeta\frac{\lambda_{1,3}}{(2\tau_{31})^{6/5}},\\
E_{32}&=\frac{7\pi}{240(2\tau_{32})^2}-\zeta^{-1}\frac{\lambda_{1,3}}{(2\tau_{32})^{6/5}},\\
E_{33}&=\frac{7\pi}{240(2\tau_{33})^2}-\zeta^{-1}\frac{\lambda_{1,3}}{(2\tau_{33})^{6/5}}.
\end{split}\label{trienergy}
\end{equation}
So for $\lambda_{1,3}<0$ (or positive coupling in the Lagrangian language) Cardy states $|(1,1)\ra,|(2,1)\ra,|(3,1)\ra$ give the degenerated minimum energy. As mentioned in \cite{Cardy}, this is an approximation of the critical Ising model. On the other hand, for $\lambda_{1,3}>0$ (or negative coupling in the Lagrangian language) Cardy states $|(2,2)\ra,|(3,2)\ra,|(3,3)\ra$ give the degenerated minimum energy. In this even $m$ case, our counting (\ref{gsd}) is consistent with the conjectured $(m-1)$-fold degeneracy in the literature \cite{H}.

\subsection{Tetracritical Ising model}
Since there is a mismatch between our proposal of ground state degeneracy and those in the literature, let us study this example in detail. Employing the Cardy's method, one finds $s=3$ minimizes the energy. So candidate Cardy states are given by
\[ |MW\ra=|(3,3)\ra,\quad|M\ra=|(4,3)\ra. \]
Surviving four TDLs act on these states as follows:
\begin{table}[H]
\begin{center}
\begin{tabular}{c|c|c}
&$|MW\ra=|(3,3)\ra$&$|M\ra=|(4,3)\ra$\\\hline
$\mcal L_{1,1}$&$|(3,3)\ra$&$|(4,3)\ra$\\\hline
$\mcal L_{2,1}$&$|(3,3)\ra+|(4,3)\ra$&$|(3,3)\ra$\\\hline
$\mcal L_{3,1}$&$|(3,3)\ra+|(4,3)\ra$&$|(3,3)\ra$\\\hline
$\mcal L_{4,1}$&$|(3,3)\ra$&$|(4,3)\ra$
\end{tabular}.
\end{center}
\end{table}
As evident from this, no other Cardy states appear. So if we believe the Cardy's method, we do not see an evidence of proposed four-fold degeneracy in the literature \cite{H}. We are not sure which ground state degeneracy is correct, $(m-1)$ or $\frac{m-1}2$.
Note that there is no invariant state because $\mcal L_{2,1}$ has quantum dimension $d_{2,1}\neq1$. Therefore, the ground state degeneracy (whether it is two-fold or four-fold) can be understood as a result of spontaneous category symmetry breaking. This example also shows that a non-trivial `representation' of BFC can be smaller than its rank. In fact, the rank four BFC is consistently realized on the two-dimensional space $\{|(3,3)\ra,|(4,3)\ra\}$. In this sense, the TQFT (whether it is correct or not) is `overconstrained.'

\end{document}